\documentclass[11pt,a4paper]{article}
\usepackage{jcappub}

\title{\boldmath Particle creation phenomenology,  Dirac sea and the induced Weyl and 
	Einstein--dilaton gravity}
\author[a]{V.\,A.\,Berezin,}
\author[a,b,1]{V.\,I.\,Dokuchaev\note{Corresponding author.}}
\author[a]{and Yu.\,N.\,Eroshenko}
\affiliation[a]{Institute for Nuclear Research, Russian Academy of Sciences, \\
	60th October Anniversary Prospect 7a, 117312 Moscow, Russia}
\affiliation[b]{Department of Elementary Particle Physics, National Research Nuclear University MEPhI (Moscow Engineering Physics Institute), Kashirskoye shosse 31, Moscow, 115409, Russia}
\emailAdd{berezin@inr.ac.ru}
\emailAdd{dokuchaev@inr.ac.ru}
\emailAdd{eroshenko@inr.ac.ru}

\abstract{We constructed the conformally invariant model for scalar particle creation induced by strong gravitational fields. Starting from the  ``usual'' hydrodynamical description of the particle motion written in the Eulerian coordinates we substituted the particle number conservation law (which enters the formalism) by ``the particle creation law'', proportional to the square of the Weyl tensor (following the famous result by Ya.\,B.\,Zel`dovich and A.\,A.\,Starobinsky). Then, demanding the conformal invariance of the whole dynamical system, we have got both the (Weyl)-conformal gravity and the Einstein--Hilbert gravity action integral with dilaton field. Thus, we obtained something like the induced gravity suggested first by A.\,D.\,Sakharov. It is shown that the resulting system is self-consistent. We considered also the vacuum equations. It is shown that, beside the ``empty vacuum'', there may exist the ``dynamical vacuum'', which is nothing more but the Dirac sea.}

\begin{document}
	\maketitle
	\flushbottom

\section{Introduction}

One of the most intriguing consequences of the quantum field theory is the phenomenon of particle creation. It is explained by the distortion of the vacuum state by the presence of  some external fields. Everybody knows that the quantum fields must be renormalized in order to produce finite physically acceptable results. In the curved space-times the situation become much more subtle. First of all, there appear two new aspects in the renormalization procedure, local and global ones. The local aspect consists in that the counter--terms, needed to compensate the divergences in one--loop quantum calculations, contain the quadratic contributions of the Riemann curvature tensor and its convolutions, the Ricci tensor and scalar curvature, which are absent in the primordial Einstein--Hilbert action. This led A.\,D.\,Sakharov to the idea that the gravitational field is not fundamental but is just the manifestation of the vacuum fluctuations of all other fields \cite{Sakh}, known nowadays as ``the induced gravity''. The global aspect is that in the curved space-times there can exist the event horizons. They will change drastically their global geometrical structure and will influence the behavior of the quantum wave functions. The remarkable example is the black hole evaporation discovered  by S.\,W.\,Hawking  \cite{Hawking74,Hawking75}. The event horizon may appear even in the locally flat space-times with conical singularity, accompanying by the analogous thermal effect what was demonstrated by W.\,G.\,Unruh \cite{Unruh}. This is connected to the non-inertial motion of the observers. In this paper we will be interested in the local aspects only

The cosmological particle creation in the framework of General Relativity  were studied extensively in 70-s of the last century by many authors \cite{Parker69,GribMam69,Zeld70,ZeldStar71,ZeldPit71,ParkerFull73,HuFullPar73,FullParHu74,FullPar74, LukashStar74,ZS77,BerKuzTk83}. Due to results of their works we know much about the structure of the counter-terms, the importance of the trace anomaly  in the particle creation processes, the rate of particle production and so on.

All the above-mentioned investigations was confined to considering the quantum scalar field on the given background metrics, namely, cosmological homogeneous, but slightly aniso\-tro\-pic, space-times.  What about the back reaction? The main obstacle in accounting for the back reaction is that the rigorous solution  of the quantum problem requires the knowledge of the boundary conditions, while the latter can be imposed only after solving the (classical) Einstein equations. Thus, we have got the ``vicious circle''.

Meanwhile, the back reaction seems very important, because not only the already created particles will change the geometry, but the very process of creation, being the pure quantum phenomenon of changing the vacuum energetic structure, should affect the classical gravitational field and may violate the well known energy dominance condition (see, e.\,g., \cite{Ber87a,Ber14}). Therefore, the back reaction influence may appear crucial in constructing the global space-time geometry.

To avoid this difficulty, we propose to describe the particle creation process phenomenologically, on the classical level, what should be rather reasonable when the gravitational field is strong enough (e.\,g., in the early universe and inside black holes). We will use the fundamental result by Ya.\,B.\,Zel`dovich and A.\,A.\,Starobinsky \cite{ZS77} that the rate of particle production is proportional the square of the Weyl tensor. It will also be shown that in our approach the conformal gravity action is actually incorporated into the formalism

The conformal gravity was invented by H.\,Weyl in 1918 \cite{Weyl}. His motivation was to construct the unified theory of two (known at the time) fundamental fields: electromagnetic and gravitational ones. Since the electromagnetic field (``identified with the Maxwell equations'') is invariant under the conformal transformations, H.\,Weyl proposed the conformal invariant Lagrangian for the gravitational field. Then, it was recognized that the Weyl's gravity allows only massless particles to exist. On this ground the theory was rejected by H.\,Weyl himself and by A. Einstein. But, nowadays, this unpleasant feature can be ``corrected'' by Braut--Englert--Higgs mechanism for the spontaneous symmetry breaking \cite{tHooft14}. The vacuum space-time with very high symmetry is a good candidate for the creation of the universe from ``nothing'' \cite{Vil82}. It can be easily verified that all the homogeneous isotropic space-times have zero Weyl tensor. In other words, these space-times are the vacuum solutions of the conformal gravity. The idea that the initial state of the universe should be conformal invariant is advocated also by R.\,Penrose \cite{Penr10,Penr14} and G.\,`t\,Hooft \cite{tHooft15}.

This paper is devoted to the detailed description of our model for particle creation in the conformal gravity. We will use, in particular, the specific formalism of conformal gravity from our previous papers \cite{bde1,bde2}. 

Throughout the paper we use the units $\hbar=c=1$ and the sign convention as in \cite{LL2}, i.\,e., the signature of the metric tensor is $g_{\mu\nu}$ is $(+,-,-,-)$, the Riemann curvature tensor is defined as 
\begin{equation}
R_{\phantom{0}\nu\lambda\sigma}^{\mu} =
\frac{\partial\Gamma_{\nu\sigma}^{\mu}}{\partial x^\lambda}
- \frac{\partial\Gamma_{\nu\lambda}^{\mu}}{\partial x^\sigma}
+ \Gamma_{\varkappa\lambda}^{\mu}\Gamma_{\nu\sigma}^{\varkappa}
- \Gamma_{\varkappa\sigma}^{\mu}\Gamma_{\nu\lambda}^{\varkappa},
\label{Riemann}
\end{equation}
while the Ricci tensor is the following convolution
\begin{equation}
R_{\nu\sigma}= R_{\phantom{0}\nu\mu\sigma}^{\mu}.
\label{Ricci}
\end{equation}
The scalar curvature $R=g^{\nu\sigma}R_{\nu\sigma}$, and $\Gamma_{\mu\nu}^{\lambda}$ are the metric connections, i.\,e., the covariant derivatives of the metric tensor are zero.

\section{Phenomenology of particle creation}

We start with construction of the hydrodynamical part of our model. In the ``classical'' hydrodynamics there exist two different sets of dynamical variables, the so called Lagrangian and Eulerian  coordinates. The first of them are comoving, i.\,e., the observer is sitting on some world-line. So, using the least action principle, one has to vary the trajectory of the (quasi)-particles. Since in such a case we cannot take into account the very processes of both creation and annihilation of particles ( i.\,e., trajectories), it is not appropriate for our purposes. Therefore, we need to use the Eulerian description, when the dynamical variables are fields, namely, the particle number density $n(x0)$ and the four-velocities. The action integral in this case is \cite{Ray} (for details see also \cite{Ber87}):
\begin{eqnarray}
S_{\rm hydro} &=& -\int\!\varepsilon(X,n)\sqrt{-g}\,dx+
\int\!\lambda_0(u^\mu u_\mu-1)\sqrt{-g}\,dx
+\int\!\lambda_1(nu^\mu)_{;\mu}\sqrt{-g}\,dx \nonumber \\
&& +\int\!\lambda_2X_{,\mu}u^\mu\sqrt{-g}\,dx,  
\label{Shydro}
\end{eqnarray}
where $\varepsilon(X,n)$ \ is the invariant energy density, $n(x)$  --- invariant particle number density,
$u^\mu(x)$ \ --- four-velocity of the particle flow,  $X(x)$ is the auxiliary dynamical variable introduced in order to avoid the identically zero vorticity of particle flow. It enters the action integral with the Lagrange multiplier $\lambda_2$, indicating the constraint $X_{,\mu}u^\mu=0$, i.\,e., $X(x)=const$ on the trajectories, thus enumerating them. The other two Lagrange multipliers, $\lambda_0(x)$ and $\lambda_1(x)$
are responsible, respectively, for the constraints  $u^\mu u_\mu=1$ (natural normalization of the four-velocities) and $(nu^\mu)_{;\mu}=0$  --- particle number conservation law. The semicolon ``;'' denotes a covariant derivative with respect to the metric $g_{\mu\nu}$.

Our aim is to incorporate into the formalism the particle ``creation law''
\begin{equation}
(nu^\mu)_{;\mu}=\Phi(inv)\neq0.
  \label{Phi}
\end{equation}
Evidently, the function $\Phi$ should depend on some invariants of the fields causing this particle creation. Here we would like to explore the fundamental result by Ya.\,B.\,Zel`dovich and A.\,A.\,Starobinsky \cite{ZS77} obtain for the cosmological particle production
\begin{equation}
(nu^\mu)_{;\mu}=\beta C^2.
  \label{beta}
\end{equation}
where $C^2$ is the square of the Weyl tensor $C^{\mu}_{\phantom{0}\nu\lambda\sigma}$ (its definition as well as some most important properties see e.\,g., in \cite{bde1,bde2}) and the coefficient $\beta$ depends on the type of particles under consideration. We will consider this ``creation law'' as our first postulate. So, the hydrodynamical part of the action integral now becomes
\begin{eqnarray}
S_{\rm hydro} &=& -\int\!\varepsilon(X,n)\sqrt{-g}\,dx
+\int\!\lambda_0(u^\mu u_\mu-1)\sqrt{-g}\,dx 
+\int\!\lambda_1\left((nu^\mu)_{;\mu}-\beta C^2\right)\sqrt{-g}\,dx
\nonumber \\
 &&+\int\!\lambda_2X_{,\mu}u^\mu\sqrt{-g}\,dx.
\label{Shydro2} 
\end{eqnarray}
Very important note. The Lagrange multiplier $\lambda_1$ is, actually, defined up to the additive constant. Indeed, let us replace $\lambda_1\rightarrow\lambda_1+\gamma_0$, $\gamma_0=const$, then
\begin{equation}
 \gamma_0\int\!\left((nu^\mu)_{;\mu}-\beta C^2\right)\sqrt{-g}\,dx \nonumber \\
=\gamma_0\int\!\left((n\sqrt{-g}u^\mu)_{,\mu}-\beta C^2\sqrt{-g}\right)\,dx  
\label{gamma0} 
\end{equation}
Due to the identity $(nu^\mu)_{;\mu}\sqrt{-g}=(n\sqrt{-g}u^\mu)_{,\mu}$, the corresponding volume integral transforms into the surface integral with no effect on the dynamical equations. In result, we are left with the same ``creation law'' as before plus the Weyl gravitational action  
\begin{equation}
S_{\rm grav}^{\rm Weyl}=-\,\gamma_0\beta\int C^2\sqrt{-g}\,dx.
\label{Weulbeta}
\end{equation}
Thus, the conformal gravity is intrinsically contained in our hydrodynamical part of the total action integral, prior to the introducing the gravitational action itself!

\section{Scalar field and conformal invariance}

By the conformal transformation we will understand the space-time dependent scaling of the metric tensor $g_{\mu\nu}$,
\begin{equation}
ds^2=g_{\mu\nu}(x)dx^\mu dx^\nu=\Omega^2\hat g_{\mu\nu}(x)dx^\mu dx^\nu
=\Omega^2(x)\hat ds^2\!.
\label{hat}
\end{equation}
The conformal invariance means 
\begin{equation}
\frac{\delta S_{\rm tot}}{\delta\Omega}=0. 
\label{inv}
\end{equation}
Therefore, we can (and will) consider the conformal factor $\Omega$ as a dynamical variable and make variations independently in $\Omega$ and in $\hat g_{\mu\nu}$ \cite{tHooft15}.

Let us $S_{\rm tot}=S_{\rm grav}+S_{\rm matter}$. By definition 
\begin{equation}
\delta S_{\rm matter}=\frac{1}{2}\int T_{\mu\nu}\sqrt{-g}\,\delta g^{\mu\nu}dx, \quad
\delta S_{\rm matter}=
\frac{1}{2}\int\hat T_{\mu\nu}\sqrt{-\hat g}\,\delta\hat g^{\mu\nu}dx,
\label{Shydro2} 
\end{equation}
where $T_{\mu\nu}(\hat T_{\mu\nu})$ is the matter energy-momentum tensor. Consider, first, the following transformation of the metric tensor
\begin{equation}
 \delta g^{\mu\nu}=-\,\frac{2}{\Omega^3}\hat g^{\mu\nu}\delta\Omega
 =-\,\frac{2}{\Omega}g^{\mu\nu}\delta\Omega.
\label{metrictrans}
\end{equation}
Suppose 
\begin{equation}
\frac{\delta S_{\rm grav}}{\delta\Omega}=0,
\label{grav}
\end{equation}
then
\begin{equation}
 0=\delta S_{\rm matter}=
 -\int T_{\mu\nu}g^{\mu\nu}\frac{\delta\Omega}{\Omega}\sqrt{-g}\,dx,
\label{deltamatter}
\end{equation}
that is, the trace of the energy-momentum tensor should be zero:
\begin{equation}
{\rm Tr}\,(T_{\mu\nu})={\rm Tr}\,(\hat T_{\mu\nu})=0.
\label{trace}
\end{equation}
If one considers the metric tensor transformation of the kind
\begin{equation}
\delta g^{\mu\nu}=\Omega^2\delta\hat g^{\mu\nu},
\label{metrictrans2}
\end{equation}
then, as can be easily seen,
\begin{equation}
\hat T_{\mu\nu}=\Omega^2T_{\mu\nu}, \quad \hat T^\mu_\nu=\Omega^4T^\mu_\nu,
\quad  \hat T^{\mu\nu}=\Omega^6T^{\mu\nu}.
\label{grav3}
\end{equation}

Let us go further on. The question arises: quanta of what kind a field are creating? The most simple choice is the scalar field. And the simplest action integral is 
\begin{equation}
S_{\rm scalar} =\int\left(\frac{1}{2}\chi^\mu \chi_\mu - \frac{1}{2}m^2\chi^2\right)\sqrt{-g}\,dx.
\label{scalatact}
\end{equation}
Here $\chi_\mu=\chi_{,\mu}$ (comma denotes the partial derivative), $\chi^\mu=g^{\mu\nu}\chi_\nu$ and $m$ is some constant with the dimension of mass. After the ``standard'' conformal transformations, namely
\begin{equation}
g^{\mu\nu}=\Omega^2\hat g^{\mu\nu},  \quad  \chi=\frac{1}{\Omega}\hat \chi, 
\label{transf}
\end{equation}
one gets
\begin{equation}
S_{\rm scalar}=\int\!\left(\frac{1}{2}\hat \chi^\mu \hat \chi_\mu
-\frac{1}{\Omega}\hat \chi_\mu \Omega^\mu \frac{1}{2}\frac{\hat \chi^2}{\Omega^2}\Omega_\mu\Omega^\mu
-\frac{1}{2}m^2\Omega^2\hat\chi^2\!\right)\sqrt{-\hat g}\,dx.
\label{scalarS} 
\end{equation}
Now indices are raising and lowering with the metric $\hat g_{\mu\nu}(\hat g^{\mu\nu})$. How to make this action conformally covariant? The recipe is well known: one should add into the Lagrangian the term $(R/12)\chi^2$, where $R$ is the scalar curvature, constructing from the metric $g_{\mu\nu}$. The result is 
\begin{eqnarray}
 S_{\rm scalar}&=&\int\!\left(\frac{1}{2}\chi^\mu \chi_\mu+\frac{R}{12}\chi^2- \frac{1}{2}m^2\chi^2\right)\sqrt{-g}\,dx \nonumber \\
 &=&\int\!\left(\frac{1}{2}\hat \chi^\mu\hat \chi_\mu+\frac{\hat R}{12}\hat\chi^2- \frac{1}{2}m^2\Omega^2\hat \chi^2\!\right)\sqrt{-\hat g}\,dx 
 -\frac{1}{2}\int\!\left(\hat \chi^2\frac{\Omega^\lambda}{\Omega}\right)_{|\lambda}
 \sqrt{-\hat g}\,dx.
 \label{scalarScov} 
\end{eqnarray}
Here the vertical line ``$|$'' denotes the covariant derivative with respect to the metric $\hat g_{\mu\nu}$. The last term can be transformed to the surface integral, it does not effect the dynamics. Remarkably enough, that started with no gravitational action  at all, we have got now both the conformal gravity (as a part of the ``creation law'') and the Einstein-Hilbert-dilaton gravity (as a part of the conformally covariant scalar field Lagrangian). 

It seems that if one puts $m=0$, everything else will be all right. But, it is not so easy, there exists a problem \cite{tHooft15}. This problem concerns the signs. With the ``correct'' sign for the kinetic term $(1/2)\chi^\mu\chi_\nu$, we have the ``wrong'' sign for the Einstein-Hilbert-dilaton part, $+(1/12)\hat R\hat\chi^2$ (with our sign convention there should be ``-'' instead of ``+''), and vice-versa. Our choice is the ``correct'' sign  for $\hat R$, i.\,e., $-(1/12)\hat R\hat\chi^2$, and the ``wrong'' sign for the kinetic term,  i.\,e., $-(1/2)\chi^\mu\chi_\nu$. This requires some explanation. First of all, we do not care about the ``correct'' sign for the kinetic term, because our scalar $\chi$ is not the ``genuine'' (i.\,e., fundamental) one. Some part of it we have already ``used'' as the created particles. The residual part can be viewed as the vacuum fluctuations that consist of virtual particles, including the conformal anomaly, which is responsible for the creation process. Moreover, the ``wrong'' sign in the kinetic term means the absence of the lower bound for the energy and allows even infinite number of the created particles (let us remember the C-field in the ``steady state'' cosmological model by F.\,Hoyle and J.\,V.\,Narlikar \cite{HoyleNar}). Besides, we are not going to consider our field $\chi$ as an independent dynamical variable. One more thing. If the scalar field $\chi$ is an independent dynamical variable, then, why it ``knows'' about the conformal transformation $g_{\mu\nu}=\Omega^2\hat g_{\mu\nu}$ and adjusts itself properly, \i.\,e., $\hat\chi=\Omega\chi$? Only, when this field is a part of it! Fortunately, in our case it is not so, and one can always choose the conformal factor, $\Omega=\varphi$, in such a way that 
\begin{equation}
\hat\chi=\frac{1}{\ell}\varphi,
\label{scalatact}
\end{equation}
where $\ell$ is some factor having dimension of length (it is introduced in order to keep the action integral dimensionless). Then, the action integral for the scalar field takes the form
\begin{equation}
S_{\rm scalar}=
-\,\frac{1}{\ell^2}\,\int\!\left(\frac{1}{2}\,\varphi^\mu\varphi_\mu
+\frac{\hat R}{12}\varphi^2
+\frac{1}{2}\,m^2\varphi^4\!\right)\!\sqrt{-\hat g}\,dx.
\label{scalarfin}
\end{equation}
There appears the self-interaction term, $\varphi^4$. It must be noted that the power $4$ in this term is only in the case of the four-dimensional space-time (it depends on the space-time dimensions). Here, two comments are in order. First, the above action is covariant under the conformal transformation, $\varphi(new)=\hat\Omega\varphi(old)$, $g_{\mu\nu}(old)=\hat\Omega^2g_{\mu\nu}(new)$, $\sqrt{- g}(old)=\hat\Omega\sqrt{-g(new)}$, what can be easily checked. Second, it is now evident, that $3m^2=\Lambda$ plays the role of the (bare) cosmological term.

To finish this Section we write down the energy-momentum tensor $T_{\mu\nu}$ for our (new) scalar field $\varphi$, obtained by varying $S_{\rm scalar}$ in $\hat g_{\mu\nu}$:
\begin{eqnarray}
\hat T_{\mu\nu}^{\rm scalar}&=&-\,\frac{1}{\ell^2}\varphi_\mu\varphi_\mu
+\frac{1}{2\ell^2}\varphi^\sigma\varphi_\sigma\hat g_{\mu\nu}
+\frac{1}{2\ell^2}m^2\varphi^4\hat g_{\mu\nu}  \\
  &&-\,\frac{1}{6\ell^2}\left(\varphi^2(\hat R_{\mu\nu}
  -\frac{1}{2}\hat g_{\mu\nu}\hat R) -2\left((\varphi\varphi_\nu)_{|\mu}
-(\varphi\varphi^\sigma)_{|\sigma}\,\hat g_{\mu\nu}\right)\!\right)\!. \nonumber
\label{scalarT}
\end{eqnarray}
Note the appearance of the second derivatives. The trace of this tensor equals
\begin{equation}
{\rm Tr}\,(\hat T_{\mu\nu}^{\rm scalar})=-\,\frac{1}{\ell^2}\left(\varphi\varphi^\sigma_{|\sigma}
-\frac{\hat R}{6}\varphi^2-2m^2\varphi^4\right)\!.
\label{scalarTr}
\end{equation}

\section{Hydrodynamics and conformal covariance}
 
Since we consider now the conformal factor $\varphi$ and transformed metric tensor $\hat g_{\mu\nu}$ as the independent dynamical variables, the above-written hydrodynamical action integral should be properly ``updated''. Let us start with analyzing the ``creation law'',
\begin{equation}
0=((nu^\mu)_{;\mu}-\beta C^2)\sqrt{-g}=((n\sqrt{-g}u^\mu)_{,\mu}-\beta C^2\sqrt{-g}).  
\label{creation}
\end{equation}
It is well known that in the four-dimensional space-time the combination $C^2\sqrt{-g}$ is invariant under conformal transformation, i.\,e., 
\begin{equation}
C^2\sqrt{-g}=\hat C^2\sqrt{-\hat g}.
\label{C2}
\end{equation}
So should be the full derivative $(nu^\mu\sqrt{-g})_{,\mu}$. The square of the interval $ds^2$ transforms as 
\begin{equation}
ds^2=g_{\mu\nu}dx^\mu dx^\nu=\varphi^2\hat g_{\mu\nu}dx^\mu dx^\nu =\varphi^2d\hat s^2,
\label{ds2}
\end{equation}
therefore, the four-velocity $u^\mu$ behaves as
\begin{equation}
u^\mu=\frac{dx^\mu}{ds}=\frac{1}{\varphi}\frac{dx^\mu}{d\hat s}=\frac{1}{\varphi} \hat u^\mu,
\label{4u}
\end{equation}
and, respectively,
\begin{equation}
 u_\mu=g_{\mu\nu}u^\nu=\varphi\hat g_{\mu\nu}\hat u^\nu= \varphi \hat u_\mu.
\label{4u}
\end{equation}
Thus
\begin{equation}
n\sqrt{-g}u^\mu=n\varphi^3\sqrt{-\hat g}\hat u^\mu=\hat n \hat u^\mu,
\label{nsqrt}
\end{equation}
where we introduced the new notation
\begin{equation}
\hat n =n\varphi^3\sqrt{-\hat g}.
\label{hatn}
\end{equation}
It is clear that in the comoving coordinate system $\hat n$  is nothing but the particle number per unit spatial coordinate volume, and, thus, the conformally invariant quantity. So, the ``creation law'' does not contain the conformal factor $\varphi$ explicitly. Therefore, the hydrodynamical part of the total action integral becomes now
\begin{eqnarray}
S_{\rm hydro} &=& 
-\int\!\varepsilon\left(X,\frac{\hat n}{\varphi^3\sqrt{-\hat g}}\right)\varphi^4\sqrt{-\hat g}\,dx 
+\int\!\lambda_0(\hat u^\mu\hat u_\mu-1)\varphi^4\sqrt{-\hat g}\,dx \nonumber \\
&&+\int\!\lambda_1\left((\hat n\hat u^\mu)_{,\mu}-\beta\hat C^2\sqrt{-\hat g}\right)\,dx
+\int\!\lambda_2X_{,\mu}\hat u^\mu\varphi^3\sqrt{-\hat g}\,dx
\end{eqnarray}
and now the hydrodynamical variables are $\hat n$, $\hat u^\mu$ and $X$. Let us write down the corresponding equations of motion 
\begin{eqnarray}
\frac{\delta S_{\rm hydro}}{\delta\hat n}&=&
-\,\frac{\partial\epsilon}{\partial n}\frac{1}{\varphi^3\sqrt{-\hat g}}
-\lambda_{1,\sigma}\hat u^\sigma=0, \\
\frac{\delta S_{\rm hydro}}{\delta\hat u^\mu}&=&
2\lambda_0\hat u_\mu\varphi^4+\lambda_2\varphi^3X_{,\mu}
-\lambda_{1,\mu}\frac{\hat n}{\sqrt{-\hat g}}=0, 
 \\
\frac{\delta S_{\rm hydro}}{\delta X}&=&
-\,\frac{\partial\epsilon}{\partial X}\varphi^4
-\frac{(\lambda_2\varphi^3\sqrt{-\hat g}\hat u^\sigma)_{,\sigma}}{\sqrt{-\hat g}}=0. 
\label{motion}
\end{eqnarray}
To these we should add, of course, the constraints that follow from variation of the action integral in Lagrange multipliers $\lambda_0$, $\lambda_1$ and $\lambda_2$:
\begin{equation}
\hat u^\sigma\hat u_\sigma=u^\sigma u_\sigma=1, \quad 
X_\sigma\hat u^\sigma=X_\sigma u^\sigma=0, \quad  
(\hat n\hat u^\mu)_{,\mu}=\beta \hat C^2\sqrt{-\hat g},
\label{constr}
\end{equation}
the last of them being equivalent to $(nu^\mu)_{;\mu}=\beta C^2$. The above equations of motion can be also written in terms of the quantities without ``hats'', namely
\begin{eqnarray}
&&-\,\frac{\partial\epsilon}{\partial n}-\lambda_{1,\sigma}u^\sigma=0, \\
&& 2\lambda_0u_\mu+\lambda_2X_{,\mu}-n\lambda_{1,\mu}=0, \\
&& -\,\frac{\partial\epsilon}{\partial X}-(\lambda_2u^\sigma)_{,\sigma}=0. 
\label{nohats}
\end{eqnarray}
It is not difficult to extract the Lagrange multiplier $\lambda_0$ from these equations. Indeed, by making the convolution of the second of the equations with the four-velocity vector $u^\mu$ and using the constraints, we get, after comparing the results with the first of the equations, that
\begin{equation}
2\lambda_0=-\,n\,\frac{\partial\epsilon}{\partial n}.
\label{lambda0}
\end{equation}
Then, introducing the pressure $p$ in the usual way, 
$p=-\epsilon+n\frac{\partial\epsilon}{\partial n}$, one obtains 
\begin{equation}
2\lambda_0=-(\epsilon+p).
\label{lambda0b}
\end{equation}
The next step is to compute the hydrodynamical part of the total energy-momentum tensor. Omitting the details, we present here the result:
\begin{eqnarray}
\hat T_{\mu\nu}^{\rm hydro}&=& -\,\frac{\hat n}{\varphi^3\sqrt{-\hat g}}
\frac{\partial\epsilon}{\partial n}\hat g_{\mu\nu}\varphi^4+\epsilon\varphi^4\hat g_{\mu\nu} 
-2\lambda_0\varphi^4\hat u^\mu\hat u^\nu
-\lambda_0(\hat u^\sigma\hat u_\sigma-1)\varphi^4\hat g_{\mu\nu}
-\lambda_2X_{,\sigma}\hat u^\sigma\varphi^3\hat g_{\mu\nu} \nonumber \\
&&-4\beta\left((\lambda_1\hat C_{\mu\sigma\nu\lambda})^{|\lambda|\sigma}
+\frac{1}{2}\lambda_1\hat C_{\mu\lambda\nu\sigma}\hat R^{\lambda\sigma}\right)\!. 
\label{Thydro}
\end{eqnarray}
or
\begin{equation}
\hat T_{\mu\nu}^{\rm hydro}= (\varepsilon+p)\varphi^4\hat u_\mu\hat u_\nu
-p\varphi^4\hat g_{\mu\nu} 
-4\beta\left((\lambda_1\hat C_{\mu\sigma\nu\lambda})^{|\lambda|\sigma}
+\frac{1}{2}\lambda_1\hat C_{\mu\lambda\nu\sigma}\hat R^{\lambda\sigma}\right)\!. 
\label{Thydro2}
\end{equation}
with the trace, equals to
\begin{equation}
 {\rm Tr}\,(T_{\mu\nu}^{\rm hydro})=(\varepsilon-3p)\varphi^4.
\label{Trhydro}
\end{equation}
The total trace equals 
\begin{equation}
{\rm Tr}\,(T_{\mu\nu}^{\rm tot})=-\,\frac{1}{\ell^2}\left(\varphi\varphi^\sigma_{|\sigma}
-\frac{\hat R}{6}\varphi^2-2m^2\varphi^4\right)+(\varepsilon-3p)\varphi^4.
\label{Trtotal}
\end{equation}
Finally, let us write the result of the variation of the total action integral in $\varphi$, which can be considered as one of the equations of motion as well as the consequence of the postulated conformal invariance. One gets
\begin{equation}
\frac{1}{\ell^2}(\varphi^\sigma_{\phantom{0}|\sigma}-\frac{1}{6}\hat R\varphi-2m^2\varphi^3)+(\varepsilon-3p)\varphi^3=0,
\label{varphieq}
\end{equation}
as it should be: ${\rm Tr}\,(T_{\mu\nu}^{\rm tot})=0$! The relation $\hat T_{\mu\nu}=\varphi T_{\mu\nu}$ can be also easily verified. This proves the self-consistency of our model. Note that in no way $\varphi$ can to be zero value, since this would lead to to the degeneracy of the whole space-time.

\section{Dirac sea and Weyl gravity}

Introducing new notations, $6l^2=8\pi G$ and $3m^2=\Lambda$, we are able to write our equations in a more familiar form (without ``hats''!)
\begin{equation}
R_{\mu\nu}-\frac{1}{2}g_{\mu\nu}R-\Lambda g_{\mu\nu}=8\pi G T_{\mu\nu}^{\rm hydro}.
\label{induced}
\end{equation}
These equations look like the ordinary Einstein equations with a cosmological constant, but now the hydrodynamical energy-momentum tensor is modified by the presence of terms originated from the ``creation law'', namely
\begin{equation}
T_{\mu\nu}^{\rm hydro}=(\epsilon+p)u_\mu u_\nu-pg_{\mu\nu}-4\beta B_{\mu\nu}[\lambda_1],
\label{Thydro3}
\end{equation}
where
\begin{equation}
B_{\mu\nu}[\lambda_1]=(\lambda_1C_{\mu\sigma\nu\lambda})^{;\lambda;\sigma}
+\frac{1}{2}\lambda_1C_{\mu\lambda\nu\sigma}R^{\lambda\sigma},
\label{Thydro2}
\end{equation}
for $\lambda_1=1$ it is just the Bach tensor $B_{\mu\nu}$. Note, that $\hat B_{\mu\nu}=\varphi^2B_{\mu\nu}$ and $\hat T_{\mu\nu}=\varphi^2T_{\mu\nu}$. It can be checked that derived equations are confomally covariant, i.\,e., if one makes the conformal transformation $\hat g_{\mu\nu}=\Omega^2\hat{\hat g}_{\mu\nu}\;\left(= g_{\mu\nu}=(\varphi^2\Omega^2)\hat{\hat g}_{\mu\nu}\right)$, then the equations written in terms of $\{(\varphi\Omega),\hat{\hat {\cal O}}\}$ look the same as for $\{\varphi,\hat {\cal O}\}$.

Let us consider the vacuum space-times. In the absence of real particles $(\epsilon=p=0)$ the energy-momentum tensor does not reveal the structure dictated by the presence of the trace anomaly. All these vacua are absolutely empty. The way out of such a situation we see in introducing yet another type of particles, but with the {\sl negative energies}. This is something like the Dirac sea. For the vacuum solutions they must compensate each other. One should not be afraid of fluctuations having negative energies above the vacuum state. Due to the self-antigravitaion they will be gone away, while  those with positive energies will undergo the usual gravitational instability and form the structures. Thus we need two (instead of one) parts of hydrodynamical action with, correspondingly, two sets of dynamical variables (labeled by ``$\pm$''). Let us write down the corresponding equations of motion 
\begin{eqnarray} \label{vect}
\left(\epsilon_{(\pm)}+p_{(\pm)}\right)u_\mu^{(\pm)}+\lambda_2^{(\pm)} X_{,\mu}^{(\pm)}
-n_{(\pm)}\lambda_{1,\mu}^{(\pm)}&=&0, \\
\frac{\partial \epsilon_{(\pm)}}{\partial X^{(\pm)}}
-\left(\lambda_2^{(\pm)} u^{{(\pm)}\sigma}\right)_{;\sigma}&=&0. 
\label{scalar}
\end{eqnarray}
In the vacuum, exactly as in the Dirac sea, from $\epsilon_+=-\epsilon_-$ and $n_+=n_-$, it follows that $p_+=-p_-$. Since there must be no energy or particle number flows in the vacuum, we get 
\begin{equation}
u^{(+)\mu}=u^{(-)\mu},
\label{upm}
\end{equation}
i.\,e., the trajectories of these two types of ``matter'' are the same. For this reason, the auxiliary variables $X^{(\pm)}$ are also the same. Therefore, the second (scalar) equation of motion (\ref{scalar}) gives us
\begin{equation}
X^{(+)}=X^{(-)} \quad \Rightarrow \quad \lambda_2^{(+)}=-\lambda_2^{(-)},
\label{Xpm}
\end{equation}
and from the first (vectorial) equation of motion (\ref{vect}) it follows that
\begin{equation}
\lambda^{(+)}_{1,\mu}=-\lambda^{(-)}_{1,\mu}.
\label{lambdap1m}
\end{equation}
The Lagrange multiplier $\lambda^{(\pm)}_1$ will enter as a sum in our vacuum equation, so
\begin{equation}
\lambda^{(+)}_{1}+\lambda^{(-)}_{1}=const.
\label{lambdap1sum}
\end{equation}
Finally, we obtain the following equation for what can be called ``the dynamical vacuum''
\begin{equation}
4\alpha_0B_{\mu\nu}+\frac{1}{16\pi G}\,G_{\mu\nu}-\frac{\Lambda}{16\pi G}\,g_{\mu\nu}=0,
\label{dynvac}
\end{equation}
where 
\begin{equation}
G_{\mu\nu}=R_{\mu\nu}-\frac{1}{2}\,g_{\mu\nu}R
\label{dynvac}
\end{equation}
is the Einstein tensor, and 
\begin{equation}
B_{\mu\nu}=C_{\mu\sigma\nu\lambda}^{\phantom{abcd};\lambda;\sigma}
+\frac{1}{2}\,C_{\mu\lambda\nu\sigma}R^{\lambda\sigma}.
\label{Thydro2}
\end{equation}
is the Bach tensor.

\section{Conclusions and Discussions}

{\sl Conclusions}
\smallskip

We have constructed the self-consistent conformally invariant phenomenological model for particle creation in the presence of strong gravitational fields. The word ``phenomenological'' means that we adopted classical description both for the created particles (hydrodynamics) and for the ``creation law''. 

This ``creation law'' enters the action integral with the corresponding Lagrange multiplier and substitute the particle number conservation law in the conventional hydrodynamics. The idea (and our hope) is that such an inclusion of the particle creation law straight into the least action formalism will cause the essential change in the structure of the energy-momentum tensor and will lead to the violation of the energy dominance condition and, thus, will take into account (to some extent) the quantum character of the particle creation process. This idea is not quite new, it was already explored by one of the authors. The new thing is combining of the method with the postulated conformal invariance of the whole theory. This allows to restrict the possible functional form of the ``creation law'' up to the square of the Weyl tensor. 

It appeared, to our surprise, that the above mentioned Lagrange multiplier can be determined only uo to an arbitrary constant. This means that the Weyl gravitational action is, actually, already incorporated into the formalism and does not need to be introduced it artificially. The local conformal invariance, taken as the fundamental symmetry, has one more important consequence. In order to make it possible to create particles we need some fields which quanta are these very particles. The simplest is the scalar field. One needs it also because it is the scalar self-interacting field that gives the masses to particles through the Brout--Englert--Higgs mechanism (which also makes the conformal gravity meaningful). If one uses the simplest (again!) form for the scalar field Lagrangian, i.\,e., ``the kinetic term $+$  the mass term'', then, in order to make it conformally covariant, it is necessary to introduce also the term proportional to the scalar curvature. Therefore, starting from hydrodynamics, needed for the description of the created particles and introducing the conformally invariant creation law plus the conformally covariant scalar field Lagrangian, we arrived at the conformal gravity theory with the Weyl Lagrangian plus the Einstein--Hilbert--dilaton gravity. This supports the idea, first discussed by A.\,D.\,Sakharov, about the  induced gravity. 

One more thing. In order to have the ``correct'' sign for the scalar curvature one has to choose the ``wrong'' sign for the kinetic term in the scalar field Lagrangian. But this causes no conceptual difficulties at all, since our scalar field is not the genuine (fundamental), it is simply the ``vacuum residual'' part of some entity, the ``above-vacuum'' part of which is already present in the form of the created particles, and its ``conformal anomalous'' part is already included into the ``creation law''. Therefore, it is very ``natural'' to identify the ``vacuum residual'' part with the conformal factor of the metric tensor. The mass term now plays the twofold role, it produces the self-interaction and the cosmological term, both initially absent. 

It is the ``vacuum residual'' part that will become the subject of our future investigations. We would like to study the possibility to have the spontaneous symmetry breaking allowing the particles to acquire the masses as well as the very appearing of the observers. The plausible result would be that the uniformly accelerated observer sees the thermal bath with the Unruh temperature as the vacuum state. Also, we would like to extend the form of our ``creation law'' by inclusion of other possible terms, like, say, the so called ``Euler characteristic density'', etc. We are also intending to investigate whether the dark energy problem could be solved using our model, without introducing any other sophisticated fields, couplings, and so on.

\medskip
{\sl Discussions}
\smallskip

1. The model presented above, is very minimalistic. The matter is not only in that we did not include into consideration the electromagnetic (abelian) and other (nonabelian) gauge fields, causing creation of the particle--antiparticle pairs with opposite charges. Here we restricted ourselves by the specific form of the ``creation law'', when the rate of particle production is proportional to the square of the Weyl tensor. The absence of other possible terms may be explained by the adopted conformal invariance principle. Indeed, $C^2\sqrt{-g}$ is conformally invariant. It is usually claimed that the latter is the only conformally invariant combination quadratic in Riemann curvature tensor in four dimensions. But, there exists yet another quadratic conformally invariant combination, namely, the so-called Hirzebruch--Pontryagin density $R^{\mu\nu\lambda\sigma}\,{^*}R_{\mu\nu\lambda\sigma}\sqrt{-g}$, where ``star'' means that ${^*}R_{\mu\nu\lambda\sigma}=\epsilon_{\mu\nu\alpha\beta}R^{\alpha\beta}_{\phantom{ab}\lambda\sigma}$. It is the total derivative and, therefore, when in the action integral, does not alter the equations of motion. In our model, however, it would enter together with the Lagrange multiplier ($\lambda_1$) and would have an influence on the whole situation. Of course, it is not a genuine scalar, but pseudoscalar. And may be , it is good, manifesting the $T$--violation in the irreversible particle creation processes.

2. The most important problem is how to organize the Braut--Englert--Higgs mechanism for generating particle's rest masses. The conventional line of reasoning is inapplicable here, because the usual (and the most convenient) solution $\varphi=0$ is impossible in our scheme, since it would mean the conformal factor vanishes and, thus, the very notion if the metrical space--time would become meaningless.

3. Let us imagine that the above problem is already solved. Then, we are able to construct the observers equipped with the clocks and other measurement devices, engines for correcting trajectories and all that. In the self-consistent theory these observers cannot be arbitrary at all. For example, the so called vacuum observers (those, ``sitting'' outside the matter distribution) should see ($=$ measure) different things, depending on their trajectories: the uniformly accelerated ones must be surrounded by the thermal bath, having the Unruh temperature. 

4. All these problems are for the future investigations.

\acknowledgments

The reported study was partially supported by RFBR, research project No. 15-02-05038\,a. The authors would like to thank Alexey Smirnov for numerous discussions and comments. One of us (V.\,B) is also grateful to Artyom Starodubtsev, Michail Smolyakov and Igor Gulamov for helpful discussions.

\end{document}